# Interfacial Potential Transduction for Diagnostics

Hyun-June Jang[1,*], Peuli Nath[1], Yuqin Wang[1], Mingoo Kim[1], Rohit Sai Kodam[1], Soobin Han[2], Sangmin Lee[3,4], Wookjin Na[3], Jihoon Kim[3], Xiaoao Shi[5,6], Jeff J. H. Kim[7], HyunKeun Joo[1], Byunghoon Ryu[8], Kiang-Teck Jerry Yeo[9,10], Seung-Jung Kee[11], Howard E. Katz[12], Junhong Chen[5,6], Youngung Seok[3,4], Yun Suk Huh[2], Dino Di Carlo[13], Hyou-Arm Joung[1,14]*

[1]Kompass Diagnostics Inc., Chicago, IL, 60651, USA
[2]Department of Biological Sciences and Bioengineering, Inha University, Incheon, 22212, Republic of Korea
[3]Department of Biotechnology and Bioengineering, Chonnam National University, Gwangju, 61186, Republic of Korea
[4]Department of Biomaterial Convergence, Chonnam National University, Gwangju 61186, Republic of Korea
[5]Pritzker School of Molecular Engineering, University of Chicago, Chicago, IL, 60637, USA
[6]Chemical Sciences and Engineering Division, Physical Sciences and Engineering Directorate, Argonne National Laboratory, Lemont, IL, 60439, USA
[7]Department of Biomedical Engineering, University of Illinois Chicago, Chicago, Illinois, 60607, USA
[8]Department of Mechanical Engineering, Inha University, Incheon, 22212, Republic of Korea
[9]Department of Pathology, The University of Chicago, Chicago, IL 60637, USA
[10]Pritzker School of Medicine, The University of Chicago, Chicago, IL 60637, USA
[11]Department of Laboratory Medicine, Chonnam National University Medical School and Hospital, Gwangju, 61186, Republic of Korea
[12]Department of Materials Science and Engineering, Johns Hopkins University, Baltimore, MD, 21218, USA
[13]Department of Bioengineering, University of California, Los Angeles, Los Angeles, CA, 90095, USA
[14]NAVO Health Co., Ltd., Gwangju, 61186, Republic of Korea

*Correspondence: june@kompassdx.com, hajoung@kompassdx.com

**Summary**

A major barrier to decentralized, near-patient diagnostics is the lack of a signal transduction modality that is both analytically precise and accessible at the point of care (POC)[1]. Optical readouts remain instrument-dependent and difficult to miniaturize[2], while compact electrochemical readouts are prone to matrix-dependent signal distortion, limiting their biomarker coverage[3]. Here, we introduce interfacial potential ($\Psi_s$) transduction as an instrument-independent electrical modality for portable, clinical-grade diagnostics across multiple diagnostic domains, enabled by a mechanistic framework that identifies key sample matrix parameters to allow robust extraction of target-specific signals from matrix interference. This framework is first demonstrated in a widely accessible lateral flow immunoassay format[4] through quantitative detection of estradiol, progesterone, and luteinizing hormone in human plasma with high correlation ($r^2 > 0.97$) with clinical analyzers. Broader applicability is further demonstrated with exceptional performance including glucose quantification for biochemical analysis with a limit of detection (LOD) of 0.92 µg/dL; HIV p24 capsid protein[5] using an immunomagnetic separation workflow (LOD = 44.8 fg/mL); and rapid hepatitis B virus[6] detection within 5 min via loop-mediated isothermal amplification for molecular diagnostics. Together, these results establish a $\Psi_s$ transduction paradigm that enables high-fidelity biomarker quantification in compact formats suitable for POC diagnostics.

In-vitro diagnostic tests are based on a transduction framework, a system that converts biological events into measurable signals[7]. The ability to accurately assess a wide range of biomarkers in compact, affordable, form factors has become increasingly essential in expanding healthcare beyond hospital walls, which calls for a transduction system that delivers high analytical performance across diverse assay formats without complex instrumentation and sophisticated transducers.[8]

For optical transduction, naked-eye tests based on simple lateral flow systems[4] are compact and affordable but are typically qualitative and without amplification have limited diagnostic capacity. Conversely, laboratory tests, where complex instrumentation and signal amplification chemistries can be deployed, are highly precise but require instrumentation[2] and specialized optical detectors[9], limiting point-of-care (POC) use.

On the other hand, electrochemical transduction has emerged as an alternative for POC diagnostics owing to its compact instrumentation in which integrated electrodes directly quantify Faradaic reactions triggered by biological events.[3] However, direct electrode–sample contact often invites interference from electroactive matrix species and parasitic electrochemical reactions that significantly degrade the signal-to-noise ratio.[3] Although analytical sensitivity has been improved through electrode surface functionalization to amplify target-specific signals[10] and microfluidic integration for sample matrix handling[3], their adoption in practical diagnostic applications remains largely confined to high-abundance biomarkers.[11]

As an alternative, electrical transduction based on interfacial potential ($\Psi_s$) modulation using field-effect transistor (FET) transducers[12] still maintains the compact form factor, but provides a principled basis for reducing background electrochemical reaction interference by electrically decoupling the sensing interface from redox reactions through a gate dielectric[13–15]. This principle has been exemplified by label-free, single-molecule-level detection[16], conceptually analogous to the matrix-robust operation of optical modalities such as surface plasmon resonance[17]. However, most implementations remain confined to proof-of-concept demonstrations[18] due to fundamental limitations of $\Psi_s$ transduction across differing sample ionic strengths and matrices arising from Debye screening[19–22] and undefined sample-matrix-derived interfacial perturbations[23], both of which are strongly dependent on the material and structural configuration of the FET device[24].

Here, we re-imagine $\Psi_s$ transduction through a standardized electrical transduction assay (ETA) that overcomes these previous instrument– and matrix–dependent constraints, reconciling analytical fidelity with amenability to miniaturization across diagnostic modalities. This framework relies on several innovations: (i) standardization of portable FET readout enabling direct extraction of $\Psi_s$ as an instrument-independent $\Psi_s$ signal, (ii) identification of sample matrix parameters governing $\Psi_s$ transduction, and (iii) control of biofluid-derived interference to enable quantitative integration into widely accessible lateral flow immunoassay (LFIA) formats. Building on our findings, we perform validation across representative biochemical analysis, immunoassay, and molecular diagnostic applications. In an LFIA implementation, ETA demonstrates quantitative detection of estradiol (E2), progesterone (P4), and luteinizing hormone (LH) in human plasma clinical specimens, with high correlation to clinical analyzers ($r^2 > 0.97$). Exceptional performance is further demonstrated in glucose quantification for biochemical analysis with limit of detection (LOD) of 0.92 µg/dL (0.051 µM), HIV 24-kDa capsid protein (HIV p24) detection under an immunomagnetic separation workflow (LOD = 44.8 fg/mL), and hepatitis B virus (HBV) DNA detection down to 35 copies within 5 min via loop-mediated isothermal amplification (LAMP). Together, this work demonstrates a $\Psi_s$ transduction paradigm that enables high-fidelity biomarker quantification in compact formats suitable for decentralized point-of-care diagnostics.

**ETA concept and design.** ETA comprises a $\Psi_s$ reporter, an electrode transducer, and an open-loop FET readout (Fig. 1a). The $\Psi_s$ reporter[25,26] (Fig. S1) is a unique component that generates charged products for analysis in proportion to the underlying analyte. The produced charged species accumulates at the interfacial surface of the working electrode (WE) under the electric field established by the gate bias ($V_G$) via the reference electrode (RE) (Fig. 1b), it modulates $\Psi_s$ in a manner largely independent of ionic strength, thereby overcoming Debye screening limitations. A key example of a $\Psi_s$ reporter is horseradish peroxidase (HRP), a widely adopted enzymatic label that generates protons and charged intermediates upon substrate turnover.[26] When conjugated to target-recognizing probes such as antibodies, nucleic acids, peptides, or aptamers, HRP enables target-specific charge generation across assays, linking molecular recognition to electrical signal transduction (Fig. S2).

The resulting charge modulates the $\Psi_s$ of the electrode transducer, which is electrically coupled to the gate of the FET in an open-loop configuration (Fig. 1b). This enables non-faradaic readout with high sensitivity, as semiconductor channels are highly sensitive to small variations in gate charge, in contrast to conventional electrochemical sensors that rely on faradaic electron transfer between electrodes.

ETA establishes a standardized transduction architecture across the assay module and the readout (Fig. 1b). For the assay module, featuring a diagnostic assay positioned between a RE and a WE, $\Psi_s$-dependent physicochemical parameters interacting with the WE such as the isoelectric point (pI), molecular weight (MW), and amphiphilicity were modulated to control matrix-derived effects in the $\Psi_s$ transduction. The assay module adopts two form factors (Fig. S3): a solution-state format compatible with established laboratory workflows (Fig. S4), and a solid-state format integrated into LFIA to enable accessible deployment (Fig. S5).

In an open-loop readout, the FET functions as part of the readout circuitry rather than as transducer; the electrode interface serves as the transducer (Fig. 1a), enabling extraction of $\Psi_s$ as a device-independent variable through universal calibration (Fig. 1b), as discussed in the following section. The open-loop FET readout, incorporating five parallel readout channels, measures five corresponding assay modules simultaneously (Fig. 1c). $\Psi_s$ transduction measured at the WE–RE pair serves as a unified $\Psi_s$ sensing modality across biochemical analyses, immunoassays, and molecular diagnostics (Fig. 1d).

**Standardization of open-loop readout.** Standardization of the open-loop readout began with a portable circuit designed to achieve semiconductor analyzer–level precision in FET characterization. The open-loop readout (Fig. S6) facilitates programmable $V_G$ sweeps via a microcontroller unit (MCU) (Fig. S7), yielding transfer curves comparable to those from semiconductor analyzers, with a maximum drain current ($I_D$) bias <0.05% across the curves (Fig. S8). Furthermore, long-term characterization demonstrates high operational stability, exhibiting < 0.05% drift over six months (Fig. S9).

Nonetheless, inherent variability in the transfer curves of individual FETs (~10% coefficient of variation (CV); Fig. S10a) gives rise to channel-dependent offsets, manifested as differences in absolute $I_D$ among five FETs measured under identical conditions (Fig. S10b, S10c). While this discrepancy reveals signal distortion arising from device-to-device variation in conventional real-time $I_D$ (Fig. S11a), obtaining the preferred device-independent metric, $\Delta V_{th}$, typically necessitates the repeated acquisition of full transfer curves using bulky analyzers (Fig. S11b), limiting portability[18].

To bridge this gap, universal calibration (Fig. 2a, S12) is introduced, converting real-time $I_D$ into equivalent potential ($V_E$) expressed in the same potential coordinate as $V_{th}$ shifts within the open-loop readout. This mapping utilizes linear regressions of $I_D$ versus $V_G$, derived from device-specific transfer curves, which are acquired during an initial calibration step using simple $V_G$ sweeps at a fixed drain voltage and stored in the MCU memory. During operation, these pre-stored regressions are applied in real time through firmware algorithms to convert $I_D$ into $V_E$. As a result, channel-dependent $I_D$ offsets collapse into consistent $V_E$ trajectories across heterogeneous FETs, reducing inter-channel variation to <0.07% CV (Fig. S10c). This universal calibration offers a generalized strategy applicable to most FET biosensors to eliminate device-to-device variability in real-time $I_D$ signals within a portable readout format, without reliance on bulky semiconductor analyzers.

Universal calibration assumes that biological event-induced $\Delta\Psi_s$ produces a rigid translation of the FET transfer curve along the $V_G$ axis without altering its shape or transconductance, with temperature identified as the primary factor that modulates the transfer curve shape. The $V_E$ across the clinically relevant operational temperature range of 4–37 °C exhibits temperature-induced variation <5% (Fig. 2b, S13). When the transfer curves are corrected using a near-linear temperature coefficient of approximately 1.9 mV/°C (Fig. S13f), the residual variation is reduced to below ~2% (Fig. 2b), demonstrating the capability to control temperature-dependent variation in the open-loop readout.

**Design rule for electrolyte-free electrodes.** To minimize the overall system footprint, electrodes in the assay module must be compact and affordable. The open-loop FET readout accommodates electrolyte-free REs such as silver chloride (AgCl) and copper, while achieving pH sensitivity comparable to that of a standard liquid-junction REs (Figs. 2c, S14) in contrast to conventional FET biosensors that rely on bulky liquid-junction RE[12], limiting scalable, low-cost implementation.

The performance of electrolyte-free REs is further evaluated using a representative $\Psi_s$ model reporter system[26] based on the HRP–hydrogen peroxide ($H_2O_2$)–chlorophenol (CP) reaction (Fig. S15). Copper and AgCl REs exhibit comparable performance and are interchangeable in solution-state electrode designs (Fig. S16). Copper is therefore selected as the standard RE material in this work due to its ease of manufacturing.

The contact area of the ITO WE is the primary determinant of baseline stability and signal reproducibility, as the interfacial impedance at the WE contributes substantially to the total impedance of the FET system (Fig. S17a). In contrast, the RE contact area has a negligible impact once sufficient electrical contact is established (Figs. S17b, S17c). Based on these results, the dimensions of the well-type electrode (Fig. S3) were determined, and a standard measurement protocol and condition—including buffer selection and substrate concentrations—for the model $\Psi_s$ reporter system was established (see Methods for details; Fig. S18). Also, $\Psi_s$ shifts remain stable for >5000 sec without observable drift (Fig. 2d); this operational stability effectively covers the typical duration required for general diagnostic detection.

**System validation and sample matrix interference.** Next, the universal calibration was validated at scale, enabling the determination of high analytical sensitivity and the identification of challenging matrix interference. Through the use of three FET groups from the same device model (CD4007) spanning high, medium, and low $V_{th}$, the calibration effectively mitigated distortions that otherwise produced large variations in the baselines and kinetics of raw real-time $I_D$ (Fig. S19). When evaluated across groups (two FETs per group) with three replicates, inter-device variability was reduced from 5.53% to 1.83% CV (Fig. S19f, S19g), with the calibrated $\Delta V_E$ closely matching the corresponding $\Delta V_{th}$ measured independently. Under this calibrated readout, HRP reactivity was resolved down to 1 ng/mL, exhibiting a linear concentration–reactivity relationship ($r^2$ = 0.990) (Fig. 2e).

ETA signaling system was evaluated directly in physiological media (Fig. 2f), including commercially sourced human serum and plasma with high ionic strength that impose an extremely short Debye length (<0.7 nm[27]). Such short Debye lengths strongly screen electrostatic signals from biomolecular charges located beyond the sensor surface, limiting the sensitivity of conventional label-free FET biosensors. Despite these constraints, $\Delta V_E$ was detected at HRP concentrations as low as 1 pg/mL. However, the signal patterns and dynamic ranges varied with matrix composition (Fig. 2f) with matrix-dependent drift (Fig. S20). Collectively, these observations necessitate a systematic investigation of matrix-derived interference parameters, including proteins and ionic species, which remain largely unexplored in the context of $\Psi_s$ transduction.

**Source of matrix interference and formulation constraints.** Matrix interference in $\Psi_s$ transduction originates primarily from highly charged species that perturb the interfacial charge environment at the ITO WE. Non-specific protein adsorption at the ITO WE surface represents the dominant source of signal variation in $\Psi_s$ transduction (Fig. 3a, S21), as evidenced by a clear pI-dependent suppression behavior observed across tested proteins (Fig. 3b). Proteins carrying a strong net positive charge induce pronounced signal attenuation, whereas proteins with low pI, in particular, lower MW (Fig. S22), tend to produce negligible HRP signal suppression, indicating a dependence of signal attenuation on protein charge characteristics and surface interactions with ITO.

While most ionic species fall into the mild interference category (Fig. 3c)—defined as a reduction in HRP reactivity by 10–20%, with reductions below 10% classified as no interference— divalent metal ions, including magnesium(II), copper(II), and iron(II), represent a notable exception. Specifically, they reduce HRP reactivity to near-background levels, a response consistent with severe disruption of HRP reaction kinetics[28–30]. In contrast, gold nanoparticles (AuNPs) commonly used in assay formulations[31] exhibit only mild interference, whereas highly charged biomolecules, including polyanionic heparin and cationic L-lysine, produce pronounced signal suppression due to strong interactions at the ITO surface.

**Materials screening for ETA control.** The results above indicate that assay formulation components—such as non-reducing disaccharides, polymers, and surfactants—introduced for stability and transport control can directly influence $\Psi_s$ transduction by altering charge distribution, hydration layers, and solid–liquid interfacial adsorption, necessitating systematic screening and control of formulation-induced interference to achieve reproducible ETA signaling.

Trehalose and sucrose showed no measurable effect on HRP reactivity, and dextran likewise exhibited negligible impact (Fig. 3d). In contrast, the anionic polymer dextran sulfate (DS) produced pronounced signal suppression. Reduced HRP reactivity observed with higher-MW polyvinylpyrrolidone (PVP) is attributed to its role as a surface-modifying polymer (Fig. S23); its ability to form hydrated interfacial layers on oxide substrates reduces protein adsorption and introduces steric shielding and passivation. By comparison, polyethylene glycol (PEG) exhibited negligible interference under identical conditions (Fig. 3d).

Surfactants exhibited heterogeneous effects on interfacial properties in the ETA signaling system (Fig. 3e, S24). Most surfactants induced substantial signal suppression, whereas a limited subset with intermediate hydrophilic–lipophile balance (HLB 13–18) exhibited minimal impact. In contrast, highly hydrophilic surfactants (HLB >18) and surfactants with higher molecular weight generally caused stronger suppression of HRP signals. Taken together, these screening results were leveraged to optimize and establish solid-state ETA measurements in an LFIA format, providing a materials-informed foundation for implementation.

**Solid-state measurements in complex matrices.** Integration of electrodes with LFIA (Fig. S4, S25) enables quantitative readout of LFIA strips, providing a clinically relevant testbed to validate ETA operation under realistic diagnostic conditions and complex biological matrices.

A dilution strategy offers a simple and practical approach for controlling matrix interference in complex biofluids, as it avoids the need for complex sample processing steps such as microfluidic manipulation. This approach is feasible only when the sensing platform exhibits sufficiently high sensitivity to compensate for analyte dilution, as demonstrated in prior work[26] by LOD in the fg/mL range. Evaluation of seven human plasma samples across E2, P4, and LH using LFIA test strips indicated that a 100-fold dilution was required to avoid measurable loss of HRP reactivity (Fig. 4a, S26).

For translation toward whole-blood testing, potential noise sources associated with cellular components were evaluated under optimized operating conditions. In particular, the influence of hemoglobin (HB)[32], a heme-containing and redox-active protein, was assessed due to its potential to disrupt HRP-mediated signal generation (Fig. 4b, Fig. S27). Under optimized assay conditions, including the standard 100× dilution, no measurable HB-induced interference was observed across the clinically relevant concentration range. Although signal attenuation could be induced at supraphysiological HB levels, LFIA readouts remained functional, confirming that HB does not represent a limiting interference factor for $\Psi_s$ transduction under the defined operating conditions.

Beyond matrix tolerance, long-term storage stability is a critical requirement for deployable diagnostic sensors. In HRP-based signaling, the limited stability of $H_2O_2$ constrains long-term reliability, which was addressed here by enzyme-mediated in situ $H_2O_2$ generation using choline oxidase (ChOx)[33]. While peak signal intensity was approximately 40% lower than that achieved with direct $H_2O_2$ addition (Fig. S28), this strategy maintained functional activity for over two months (Fig. 4c, S29).

**Applicability of the ETA.** Fertility hormones, including E2, P4, and LH, require frequent and time-resolved monitoring in infertility care[34], motivating accessibility-oriented, real-sample assay design in the diagnostic sector. Glucose was chosen as well-established targets in the biochemical analysis sector.[35] To demonstrate compatibility of ETA with established clinical laboratory workflows, IMS-based HIV p24 detection was selected as a representative immunoassay application, given the clinical necessity for early detection and high analytical sensitivity.[5] Additionally, LAMP-based HBV was chosen as a molecular diagnostic target, as it remains the most prevalent cause of chronic viral hepatitis worldwide.[36]

Using optimized ETA components (Fig. S30), the LFIA from human plasma samples while maintaining strong agreement with a clinical laboratory analyzer (Cobas e801) for E2, P4, and LH, with $r^2$ = 0.974, 0.975, and 0.981, respectively (Fig. 4d–f). Collectively, these results demonstrate that ETA delivers clinical-grade analytical accuracy within an accessibility-oriented diagnostic format.

The analytical sensitivity for glucose detection within the ETA platform was assessed (Fig. 4d). By leveraging glucose oxidation, where glucose oxidase (GOx) generates $H_2O_2$ (Fig. S31), the system achieved a LOD of 0.92 µg/dL (0.051 µM) (Fig. 4g). This performance extends the detection capability well below the lower threshold of conventional glucose meters (Table S1).

An IMS assay for HIV p24 antigen detection (Fig. S32) was characterized using tetramethylbenzidine (TMB), the most widely utilized substrate in HRP-based enzyme-linked immunosorbent assay[37] and higher sensitivity than CP which provides an 8.93-fold higher reactivity compared to CP (Fig. S33), enabling concentration-dependent responses with an ultralow LOD of 44.8 fg/mL (Fig. 4h, S34).

The ETA platform was further configured for LAMP, following the same sample preparation workflow as conventional optical LAMP[38] while enabling parallel electrical readout (Fig. S35). This array-based format supports multiplexed, real-time monitoring of LAMP reactions. ETA achieved clear copy-number–dependent signal separation within 5 min, including detection at 35 copies of HBV DNA (Fig. 4i, S36).

ETA-optimized LAMP formulations differ from conventional fluorescence-optimized LAMP conditions[39] (Fig. S37), reflecting design rules tailored for charge-based electrical readout rather than optical detection. To minimize charge-derived interference, ionic species were reduced during formulation optimization (Fig. S38a). In addition, surface-protective components identified in Fig. 3d were incorporated to stabilize the ITO surface during LAMP process (Fig. S38b). Finally, reagents that are highly charged and commonly used to enhance amplification efficiency in fluorescence-based LAMP assays[40]—including betaine, ammonium sulfate (($NH_4)_2SO_4$), potassium chloride (KCl), deoxyribonucleotide triphosphates (dNTPs), and magnesium sulfate ($MgSO_4$), and primers—were deliberately reduced (Fig. S38c).

**Discussion**

The gap between rigorous diagnostics and deployable formats stems from the inherent limitations of conventional signal transduction—primarily optical and electrochemical modalities—across diverse measurement contexts. To address this long-standing limitation, ETA establishes a standardized $\Psi_s$ transduction framework that transforms FET-based biosensing from proof-of-concept demonstrations into a translational diagnostic modality with state-of-the-art LOD in a portable format (Table S1), providing a viable alternative to optical and electrochemical readouts.

This capability arises from ETA's strong tolerance to sample matrix effects, enabled by its non-Faradaic FET sensing architecture, which allows matrix-dependent parameters governing $\Psi_s$ transduction to be identified and actively controlled. Thus, ETA operates reliably in an LFIA format without stringent sample processing, requiring only simple dilution, highlighting its accessibility and applicability across diverse biofluids. This advantage becomes particularly pronounced in hemolytic samples, where HB, a common interferent in optical and electrochemical assays[41], has minimal influence on ETA signal integrity. Consequently, stringent sample preparation to remove HB is unnecessary, minimizing instrumentation requirements and enabling direct whole-blood testing in POC settings.

Efforts to standardize the readout of $\Psi_s$ in FET biosensing toward translational applications have remained largely unexplored. First, the influence of complex sample matrices and operational variability has received limited attention, as most FET biosensing studies have focused on label-free detection evaluated under controlled conditions due to Debye screening limitations. Second, although $\Psi_s$ is fundamentally a potential, it is typically transduced into current signals in conventional portable FET biosensing systems, making the measurement inherently dependent on device-specific characteristics such as semiconductor material, structure, and operating conditions. We address these limitations by decoupling transduction from the FET and employing it as a readout system, enabling direct extraction of $\Psi_s$ in a portable format while allowing control over key parameters of complex sample matrices.

For broader adoption in in vitro diagnostics, ETA must be compatible with existing assays. Integration with an IMS workflow shows that ETA replaces only the downstream transduction step while preserving the assay workflow, enabling scalable integration across diverse diagnostic platforms.

Understanding how assay components influence $\Psi_s$ in ETA—including the roles of MW, pI, HLB, steric effects, and interactions among components—remains a central challenge for further maturation of the platform. Continued development of enzyme substrates and conjugates engineered for $\Psi_s$ transduction provides a clear route to improved performance, including wider signal windows and long-term storage stability compatible with dry-chemistry formats. In parallel, deeper insight into how charged reaction products modulate interfacial impedance, coupled to electrode design and FET input impedance, will be critical for improving signal fidelity and reproducibility.

Most bioanalytical assays are optimized for optical readouts, which makes it challenging to relate optical signals to electrical interfacial responses and leaves interfacial behavior primarily characterized empirically. This limitation is evident in our LAMP experiments: reducing primers, nucleotides, ions and other charged components—typically selected to optimize optical performance—produced electrical responses that diverged from optical signaling, indicating that optical readout does not reliably report interfacial electrical dynamics. Further progress in ETA will therefore require complementary analytical methods that can directly probe interfacial responses.

In conclusion, this work establishes $\Psi_s$ as a physical signaling coordinate for biosensing, redefining diagnostic transduction beyond optical and electrochemical readout and unifying biochemical analysis, immunoassay, and molecular diagnostics within a single $\Psi_s$ framework, enabling analytical rigor to be retained in deployable diagnostic formats.

**Methods**

**Open-loop FET readout system.** The open-loop reader (Fig.1c) comprises five independent sensing channels, each incorporating a discrete FET (CD4007) and a dedicated operational amplifier (op-amp). The gate of each FET is electrically coupled to an ITO/PET (Sigma-Aldrich, 639303) WE. $V_G$ is generated by digital-to-analog converters (DACs) and applied to the RE, while $V_D$ is delivered independently to each FET channel (Fig. S7). The precision of the applied voltages, generated by firmware-controlled DACs via the MCU, were verified using a digital multimeter (Fig. S8c).

A MCU running embedded firmware controls bias generation, timing, and data acquisition. The reader device is powered via a USB Type-C cable connected to a laptop computer, which also provides for data transfer and device control. On-board power management circuitry generates regulated supply voltages for digital logic and higher-voltage rails for $V_G$ and $V_D$ biasing (see Fig. S7 for details).

To measure the intrinsic FET transfer characteristics, the RE and WE terminals of each channel were electrically shorted. The $V_D$ was fixed at 3.3 V, while the $V_G$ was swept from 0 to 3.3 V in 0.05V increments and recorded in the MCU memory. $I_D$ through a series sense resistor (3 kΩ) connected at the drain, generating a voltage drop across the resistor. An op-amp measures the voltage across the resistor terminals, which is digitized by an analog-to-digital converter (ADC). In firmware, the digitized voltage is converted to the corresponding $I_D$ using the known resistance value.

Transfer characteristics measured by the open-loop reader (Fig. S8) were benchmarked against with a semiconductor parameter analyzer (Keithley 4200A). For direct comparison, a 3 kΩ drain resistor was connected, and the resulting signal was read out using the Keithley 4200A. All measurements were performed under identical $V_D$ conditions.

**Data display.** Under the operating conditions of the open-loop reader, real-time $I_D$ values are acquired at fixed $V_G$ of 1.5 V and $V_D$ of 3.3 V and calibrated to equivalent $V_E$ via a linear regression–based interpolation algorithm implemented in firmware (Fig. S12). Both $I_D$ and $V_E$ data were recorded at 1 sec interval and transmitted to a laptop through a USB Type-C connection and visualized using custom ADC readout software.

**Temperature response evaluation.** Temperature response evaluation (Fig. 2b, S13) was performed by placing the open-loop reader in controlled thermal environments with digital thermometer while keeping the laptop computer external. Low-temperature measurements were conducted in a laboratory refrigerator set to 4 °C, room-temperature measurements were performed under laboratory-controlled ambient conditions, and high-temperature measurements were carried out in a temperature-controlled dry oven at 37 °C.

**Design and fabrication of the well-type electrode.** The pH sensitivity of the ITO WE paired with different RE materials (Fig. 2c) was evaluated by dispensing 20 µL of standard pH buffer solutions between the ITO WE and each RE material, as shown in Fig. S14. The results were compared with those obtained using a standard liquid-junction Ag/AgCl RE (Gamry, 930-00015). Preliminary electrode screening (Fig. S16) was performed using a custom-designed V-shaped electrode (Fig. S15), in which the RE section was replaced while maintaining the same ITO material for the WE. The effects of RE and WE contact areas were evaluated by adjusting the spacer of the V-shaped electrode, as illustrated in Fig. S15. Buffer screening (Fig. S18c) was performed using a V-shaped electrode by varying only the buffer type while following the standard measurement protocol described below. The detailed design and fabrication process of the well-type electrode is described in Fig. S3.

**Standard measurement protocol.** A 40 µL aliquot of the reaction mixture containing HRP (Toyobo, PEO-301) and CP (Sigma-Aldrich, 185787) in PBS was introduced into the well-type electrode and monitored for 300 sec to establish a stable baseline. The enzymatic reaction was initiated by the addition of 40 µL of $H_2O_2$, and the signal was subsequently recorded for an additional 800 sec. Reactivity was quantified from the maximum slope of the recorded kinetic curve ($\Delta V_E$/sec) following reaction initiation.

**Standard measurement condition.** The optimal substrate concentrations of CP and $H_2O_2$ were established at 3 mM and 1 mM, respectively in PBS (Fig. S18). Stock solutions of 1 M CP prepared in isopropyl alcohol and 30% $H_2O_2$ (w/v) in $H_2O$ (Sigma-Aldrich, H1009) were diluted in PBS to achieve the desired final concentrations. HRP concentrations were varied as specified in each result section.

Unless otherwise noted, all measurements were conducted using the well-type electrode under the standard measurement protocol and conditions described above. The concentrations of HRP or HRP-labeled species (i.e. antibody–HRP–AuNP conjugate) and other reaction components were adjusted depending on the experimental design.

**Evaluation of electrode response stability, analytical sensitivity, and matrix effects.** The response stability and analytical sensitivity were evaluated under the standard measurement protocol and condition above with varied HRP concentrations. To assess response stability (Fig. 2d), HRP at concentrations of 20 and 100 ng/mL was continuously monitored for 5000 sec. Signal resolution was evaluated by serial dilution of HRP in PBS over a concentration range of 0–20 ng/mL at 1 ng/mL intervals (Fig. 2e).

The influence of sample matrix was examined under the standard measurement protocol and condition by replacing PBS with control human serum or clinical plasma (Fig. 2f). HRP was prepared at final concentrations ranging from 1 pg/ mL to 100 µg/mL by serial dilution directly into each matrix. Control serum experiments were conducted using commercially sourced human serum (Sigma-Aldrich, H4522), and clinical plasma samples were randomly selected from an internal cohort. $V_E$ values were extracted at a fixed time point of 600 sec to demonstrate matrix-dependent variation in $V_E$ level in Fig. 2f. All measurements were performed in replicate, and mean $V_E$ values were used for comparison across different matrices. Protein pI and MW used for analysis were obtained from vendor-provided specifications. Proteins lacking vendor-reported pI and MW information were excluded from the analysis.

**Screening of formulation components affecting interfacial potential response.** All experiments (Fig. 3) were performed using the standard measurement protocol described above, except where additional components were introduced in PBS buffer at class-specific fixed concentrations, as specified below for each experiment, while all other assay conditions were held constant. The final HRP concentration was fixed at 20 ng/mL for all measurements. To assess factors affecting the signal response of the $\Psi_s$ transduction system, four classes of additives were systematically screened: proteins (n = 15), ionic and molecular interferents (n = 10), non-reducing sugars and biocompatible polymers (n = 11), and surfactants (n = 38) (Fig. 3). Proteins spanning a broad range of pI were selected, while surfactants covering a wide range of HLB values were included. Ionic species and molecular interferents, as well as sugars and polymers commonly used in diagnostic formulations, were also evaluated. Reactivity values were normalized to the corresponding PBS-only control measured under identical conditions. Detailed information on all evaluated components, including vendor sources, physicochemical properties, and concentrations, is provided in Tables S2 and S3. HLB of surfactants used for analysis were obtained from vendor-provided specifications. Surfactants lacking vendor-reported HLB information were excluded from the analysis.

**Fabrication of standard LFIA strips, electrode integration, and assay operation.** Standard LFIA strips consisted of a sample pad (Cytiva, CF5, 2 mm × 22 mm), a NC membrane laminated on a backing card (Sigma-Aldrich, HF135MC100), and an absorbent pad (Cytiva, CF5, 2 mm × 12 mm). The NC membrane contained both capture and signal zones, with the capture zone functionalized with BSA conjugated antigens (E2 or P4) or LH capture antibody, depending on the assay configuration (Fig. S25). Reagents were dispensed onto the NC membrane using an automated dispensing system (ClaremontBio) at a dispensing rate of 0.25 µL/mm. ChOx (Toyobo, Cho-301) solution (1% w/v in PBS) was dispensed to define the signal zone at a position 7.5 mm from the upstream edge of the NC membrane. The capture zone was patterned 7.5 mm downstream from the signal zone at a concentration of 0.5 mg/mL prepared in 2% (w/v) trehalose in PBS. Following reagent dispensing, membranes were dried at 37 °C for 15 min and stored in a dry cabinet (<5% relative humidity) for 24 h. The sample pad and absorbent pad were laminated onto the NC membrane with a 2.5 mm overlap to ensure continuous capillary flow, forming an assembled card. The assembled cards were cut into 2 mm × 30 mm strips using an automated guillotine cutter (Zeta Corporation). Individual strips were affixed onto a strip-mounted jig (Fig. S4c) using double-sided adhesive tape.

For electrode integration, ITO WE and copper RE were cut to identical dimensions (2 mm × 30 mm) and mounted onto a dedicated electrode alignment jig using double-sided adhesive tape. The WE was aligned above the signal zone, while the RE was positioned at the upstream edge of the sample pad (Fig. S4). The electrode-mounted jig and the strip-mounted jig were mechanically coupled using clamps to maintain fixed electrode positioning and contact height. The distal ends of the electrodes extended beyond the jig housing to allow electrical connection to the reader via alligator clips.

For assay operation, HRP-labeled detection antibodies conjugated to 15 nm AuNPs (Ted Pella, 15704-1) were prepared as described in our previous work[26]. The HRP labeling kit (Dojindo, LK11) was used for antibody-HRP conjugation followed vender's protocol.

To validate the in-situ $H_2O_2$ generation system (Fig. S28), solution-based testing was performed first using a well-type electrode with a modified standard measurement protocol. A 40 µL mixture containing HRP, CP, and CC (Sigma-Aldrich, C1879) was injected, followed by injection of the ChOx solution. To determine the optimal conditions for in situ H$_2$O$_2$ generation in the LFIA platform (Fig. S29), strips lacking an immunoassay zone were tested using 50 mOD E2–antibody–HRP AuNP conjugates to retain HRP signaling. To minimize potential interference in HRP signaling, a basic 1% BSA running buffer without surfactant was used. Detailed running buffer composition is provided in Fig. S29. For all LFIA tests, 150 μL of running buffer was loaded, and strips were positioned at a 45° angle to prevent overflow.

To optimize surfactant components affecting immunoassay efficiency in the LFIA system (Fig. S30), running buffers composed of 1% BSA, 3 mM CP, and 15 mM CC, each supplemented with 0.25% (w/v) of a candidate surfactant, were screened using 80 pg/mL spiked E2 antigen or antigen-free controls, together with 50 mOD E2–antibody–HRP AuNP conjugates. Among the tested surfactants, 0.25% (w/v) Pluronic L64 showed the highest immunoassay performance and was incorporated into the standard running buffer which consisted of 1% BSA, 3 mM CP, 15 mM CC, and 0.25% (w/v) Pluronic L64.

**Evaluation of ETA performance in LFIA using clinical sample matrices.** For dilution studies (Fig. 4a), clinical plasma samples were randomly selected from an internal cohort. Plasma samples were diluted in a range from 10× and 200× in standard running buffer lacking 0.25% (w/v) Pluronic L64 to assess dilution-dependent matrix interference only (Fig. S26). To maintain identical antigen input across all conditions, target antigens were spiked such that the total antigen amount remained constant across dilution ratios. Final antigen concentrations were 80 pg/mL for E2, 2 ng/mL for P4, and 0.1 mIU/mL for LH, respectively. Each mixture was then loaded into the corresponding LFIA cartridge for E2, P4, or LH, and signals were compared with those obtained from running buffer–only controls containing the same total amount of antigen.

The HB interference (Fig. 4b, S27) was evaluated by spiking HB into plasma samples at defined concentrations while maintaining constant antigen input. HB stock solutions were prepared from pooled human red blood cells (Innovative Research, IWB3LYIGM) by hypotonic lysis, followed by centrifugation to remove cell debris. The clarified hemolysate was serially diluted and spiked into plasma samples prior to analysis. The plasma mixture was further diluted 100× in standard running buffer lacking 0.25% (w/v) Pluronic L64, supplemented with 50 mOD of E2–antibody–HRP–AuNP conjugates, and then loaded into the E2 LFIA cartridge. Reactivity was normalized to the HB-free PBS control.

Long-term stability of the signal zone was evaluated independently using LFIA strips containing only the signal zone (Fig. 4c). ChOx patterned strips were stored under dry conditions for up to 60 days in a dry cabinet. Signal responses were measured using 50 mOD E2–antibody–HRP–AuNP conjugate, and reactivity at each time point was compared with that obtained from freshly prepared strips (day 0) to quantify signal retention over time.

**Clinical plasma evaluation of fertility hormones using LFIA.** Clinical plasma samples for fertility hormone analysis (E2, P4, LH) were obtained from the University of Chicago Medical Center under an Institutional Review Board (IRB)–approved protocol. Hormone concentrations were first quantified using a Cobas e801 analyzer (Roche Diagnostics) as the reference method. Prior to LFIA measurement, plasma samples were diluted 100-fold in standard running buffer and supplemented with 50 mOD of the corresponding conjugate for E2, P4, or LH. The mixture was loaded onto the test strip immediately after mixing. Measurements of E2, P4, and LH (Fig. 4d-f) were then performed following the LFIA operating procedure described above. Unless otherwise noted, reported analyte concentrations correspond to the original (pre-dilution) plasma concentrations.

**Glucose measurement.** Glucose measurements (Fig. 4g) were performed with minor modifications to the standard measurement protocol. A 40 µL mixture containing 0.25 mg/mL glucose oxidase (GOx) (Toyobo, GLO-101), 0.5 mg/mL HRP, and 3 mM CP prepared in PBS was injected into the well-type electrode as a control (Fig. S31). Separately, the same mixture was prepared with the addition of 2% BSA and 0.05% Tween-20 to improve signal stability. D-(+)-Glucose (Sigma-Aldrich, G8270) standards prepared in PBS were used to initiate the enzymatic reaction and were introduced (Fig. S31). Quantitative analysis was performed using the $\Delta V_E$, defined as the voltage difference measured within the first 0–20 sec after reaction initiation.

**IMS immunoassay for p24 detection.** Magnetic beads (Dynabeads™ M-270 Carboxylic Acid; Thermo Fisher Scientific, 14305D) were functionalized with anti-p24 capture antibodies according to the manufacturer's protocol. Detection antibody–HRP conjugates and AuNP conjugates were prepared following the same procedure used for the LFIA test. All antibodies and antigens were obtained from Sinobio and Fapon. The overall IMS assay workflow is illustrated in Fig. S32. Briefly, antibody-functionalized magnetic beads, AuNP-based detection conjugates, and p24 antigen were mixed and incubated simultaneously in PBS. After 15 min of incubation, the beads were magnetically separated using a DynaMag™-2 magnet (Fisher Scientific, 12-321-D) and washed with PBST to remove unbound species. The washing step was repeated for a total of four cycles, followed by final resuspension in PBS prior to signal measurement. Following completion of the IMS process, the bead suspension was transferred to the well-type electrode. A modified standard measurement protocol was used for signal acquisition. A stable baseline was recorded for 5 min, after which the enzymatic reaction was initiated by addition of a TMB/$H_2O_2$ substrate solution (Pierce™ TMB Substrate Kit; Thermo Fisher Scientific, 34021) and monitored for 20 min. Quantitative analysis was performed using the endpoint $\Delta V_E$. A final TMB concentration of 40% (v/v) was selected for IMS-based measurements (Fig. S33).

**LAMP for HBV DNA measurement on the ETA Platform.** A well-type electrode was modified with extra chamber and used for this test (Fig. S35). LAMP was implemented on the ETA for monitoring of nucleic acid amplification process under isothermal conditions. The formulation of LAMP reaction mixtures was modified for the $\Psi_s$ transduction system (Fig. S37). The modified formulation was established through stepwise optimization of reaction components to minimize baseline drift and enhance the $\Psi_s$ transduction signal (Fig. S38). Clinical plasma samples positive for HBV were obtained from Chonnam National University Hospital (CNUH). The requirement for informed consent was waived by the IRB, and the study protocol was approved by the IRB of CNUH (CNUH No. 2025-167).

HBV DNA copy numbers were quantified using real-time quantitative PCR (RT-qPCR) on the m2000 RealTime System (Abbott) according to standard clinical protocols. Viral double-stranded DNA (dsDNA) was extracted using the QIAamp DNA extraction kit (Qiagen) following the manufacturer's instructions. HBV DNA templates at defined copy numbers were added to the reaction mixture and loaded into sealed well-type electrode (Fig. S35). A total reaction volume of 70 μL was used. The device was incubated at 65 °C in dry oven to initiate isothermal amplification. $\Delta V_E$ associated with DNA amplification monitored for up to 600 sec. Raw signals were normalized to a blank reaction lacking template DNA (Fig. S36), and reaction rates were calculated as the slope of $\Delta V_E$ over defined time windows (1–10 min), enabling copy-number-dependent discrimination of amplification kinetics (Fig. 4f). To verify that amplification under ETA conditions corresponded to LAMP, reference fluorescence-based LAMP measurements and agarose gel electrophoresis were performed using conventional reaction formulations, as shown in Fig. S37.

**Acknowledgements**

We thank Prof. Aydogan Ozcan for valuable discussions and feedback on the manuscript. This work was supported by the National Science Foundation (NSF; Grant No. 2507280).

**Author contributions**

H.-J.J. and H.-A.J. conceived and led the study. H.E.K. and D.D.C. contributed to the conceptual design of ETA. M.K., P.N., Y.W., and R.S.K. designed and performed the LFIA strip experiments. S.H. and Y.S.H. developed the IMS-integrated ETA system. X.S., H.J., and J.C. designed the solution-state measurement platform and conducted screening studies. J.J.H.K. and B.R. designed the open-loop readout circuit. S.L., W.N., J.K., and Y.S. designed and performed the LAMP experiments. K.-T.J.Y. and S.-J.K. collected and curated clinical samples for hormone and HBV analyses, respectively.

**Competing interest**

H.-J.J., H.-A.J., and M.K. are co-founders of Kompass Diagnostics. D.D.C. serves as an advisory board member of Kompass Diagnostics. The authors have a pending patent application related to the technology described in this work.

**Additional information**

The online version contains supplementary material available at

**Correspondence and requests for materials** should be addressed to H.-J.J., or H.-A.J..

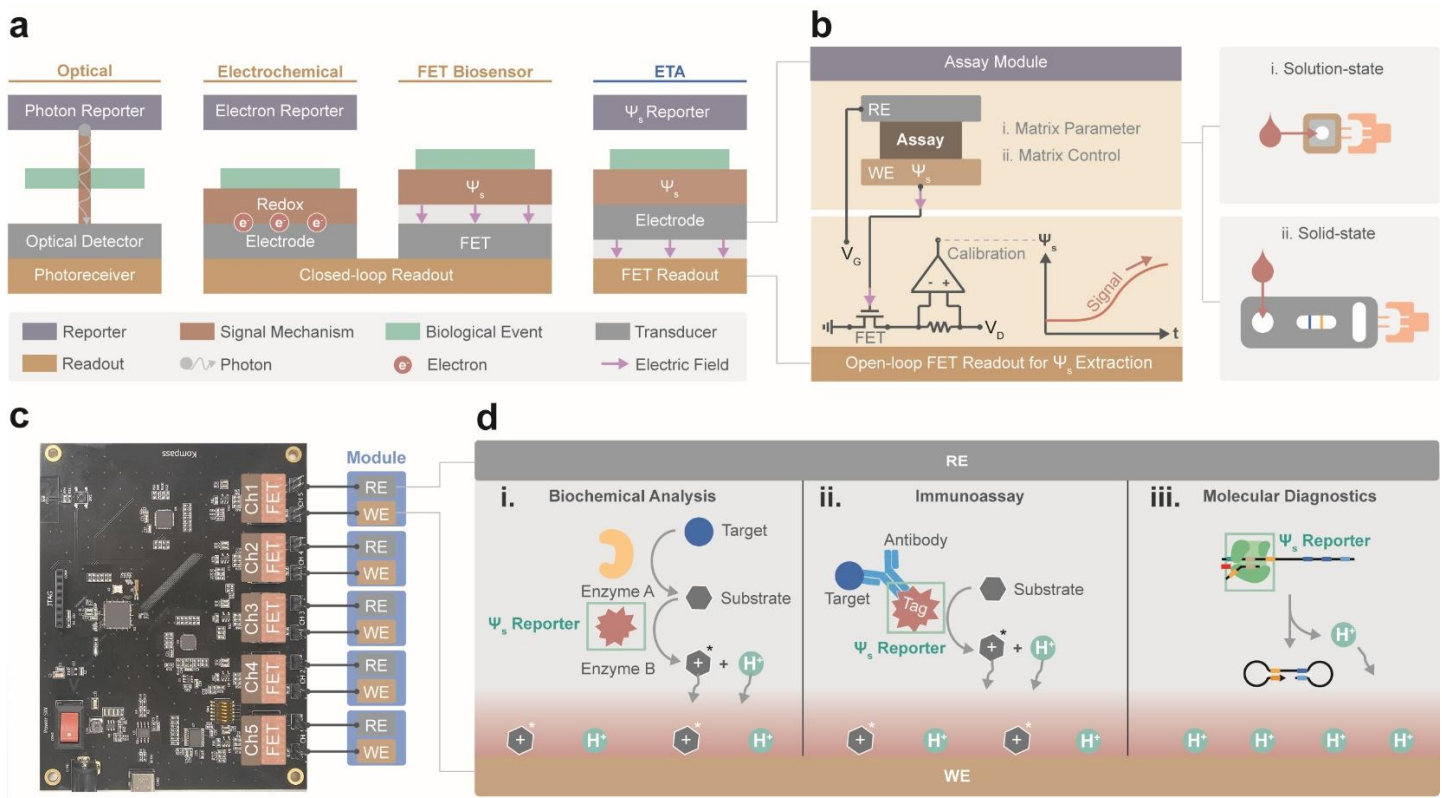

**Figure 1. ETA concept and design.** (a) Comparative biosensing framework summarizing transduction modalities, physical signal carriers, transducers, and readout schemes across optical, electrochemical, FET biosensors, and ETA. (b) Standardization framework of ETA, comprising establishment of $\Psi_s$ as an independent signaling variable, identification of sample-matrix parameters governing $\Psi_s$ transduction, and control of matrix-derived interference. Formats of assay module unit, supporting both solution-state and solid-state measurements through an open-loop FET readout. (c) Photograph of the portable five-channel FET reader integrated with modular electrode units. (d) Unified $\Psi_s$ transduction principle and its assay design space. Across assay classes, specific biological events generate charged products, modulating the $\Psi_s$ at the WE. This principle is applicable to (i) biochemical analysis via enzyme-catalyzed reactions, (ii) immunoassays via HRP-tagged antibody reactions, and (iii) molecular diagnostics via proton and/or charged product release during nucleic acid amplification cycles.

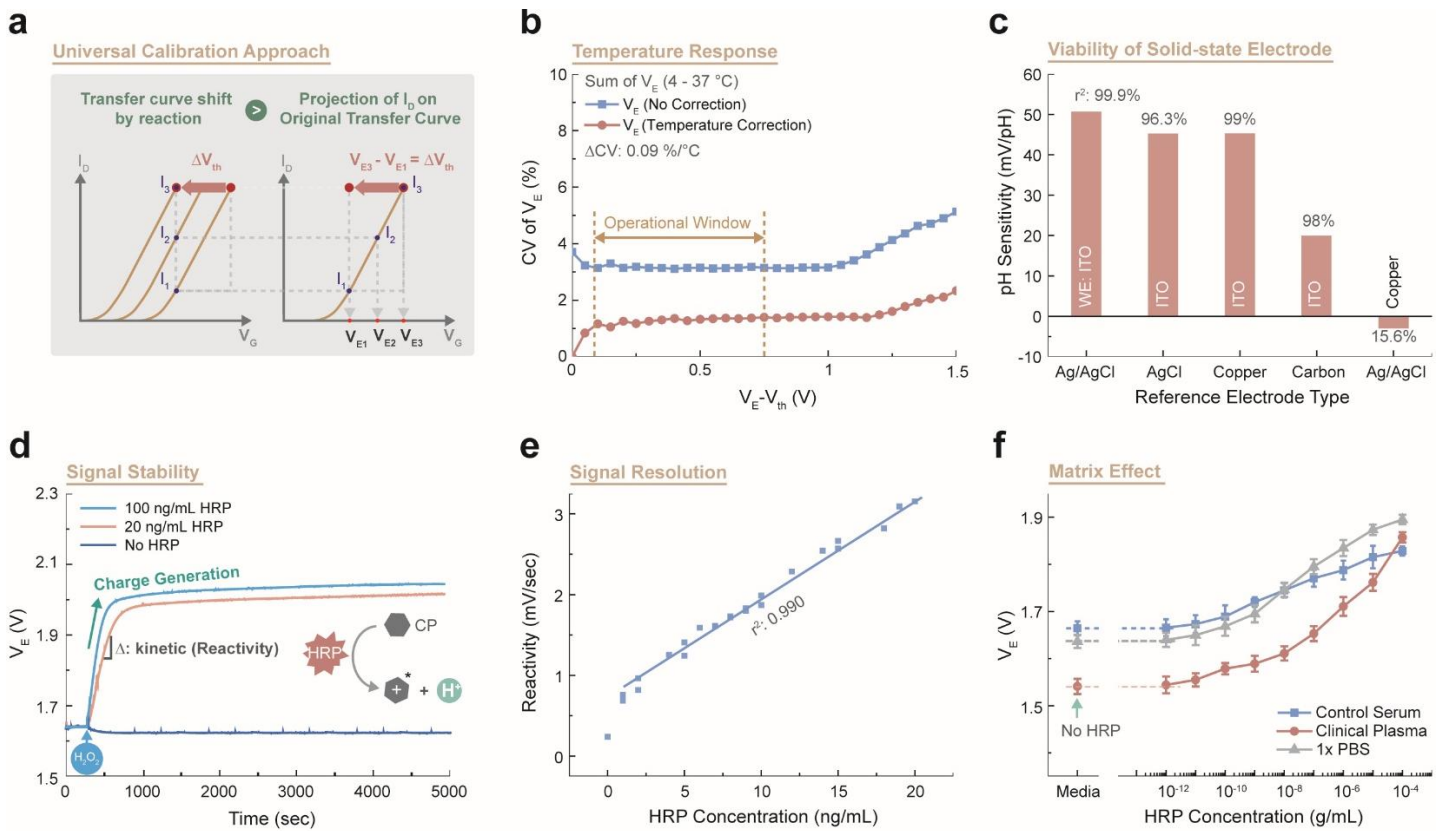

**Figure 2. Readout standardization and sensing system validation.** (a) Schematic of the universal calibration approach. Biological event-induced parallel shifts of the transfer curve are converted into $V_E$ shifts by projecting real-time $I_D$ values onto the prerecorded original transfer curve. (b) Temperature response of the calibrated $V_E$. CV of $V_E$, derived from the original transfer curve, is evaluated over 4–37 °C within the above-threshold region and plotted as a function of $V_E$-$V_{th}$. CV remains below ~5% and is further reduced to below ~2% after temperature correction within the operational window. (c) pH sensitivity of various electrolyte-free RE–WE pairings, including Ag/AgCl–ITO, AgCl–ITO, copper–ITO, carbon–ITO, and Ag/AgCl–copper. Ag/AgCl–ITO and AgCl–ITO exhibit the highest sensitivity (>96%). (d) Representative real-time $V_E$ traces of the HRP-catalyzed reaction, showing stable trajectories without noticeable drift over >5000 sec. (e) Signal resolution of the HRP-catalyzed reaction ($r^2$ = 0.99). (f) $V_E$ responses for HRP-catalyzed reaction measured in PBS, commercial control serum, and clinical plasma show consistent concentration-dependent trends for each condition, with matrix-dependent offsets in absolute response magnitude.

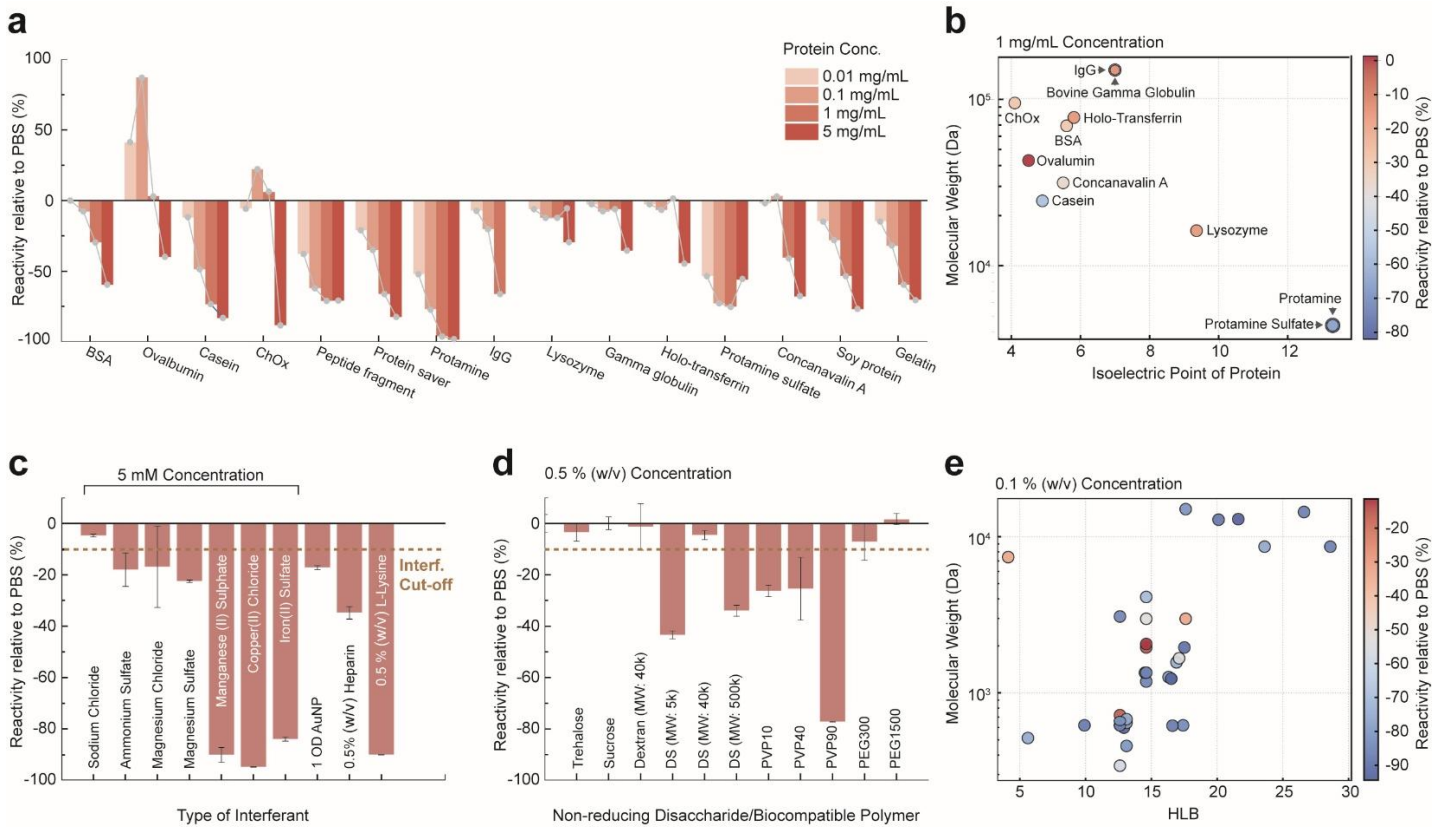

**Figure 3. Charge-driven interferences and formulation constraints.** (a) Reactivity relative to the no-protein PBS control measured in the presence of various proteins at concentrations of 0.01, 0.1, 1, and 5 mg/mL. (b) Reactivity relative to the no-protein PBS control measured in the presence of individual proteins, plotted as a function of pI and MW. (c) Reactivity changes induced by inorganic salts, metal ions, and strongly charged ionic species (including sulfate-containing polyanions and small charged molecules) at 5 mM concentration, normalized to the PBS-only condition. (d) Reactivity changes induced by common formulation stabilizers at 0.5% (w/v), normalized to the no-additive PBS control. The dashed line in (c-d) denotes the interference cutoff defined in this study as a 10% deviation in reactivity. (e) Reactivity relative to the no-surfactant PBS control as a function of surfactant HLB and MW.

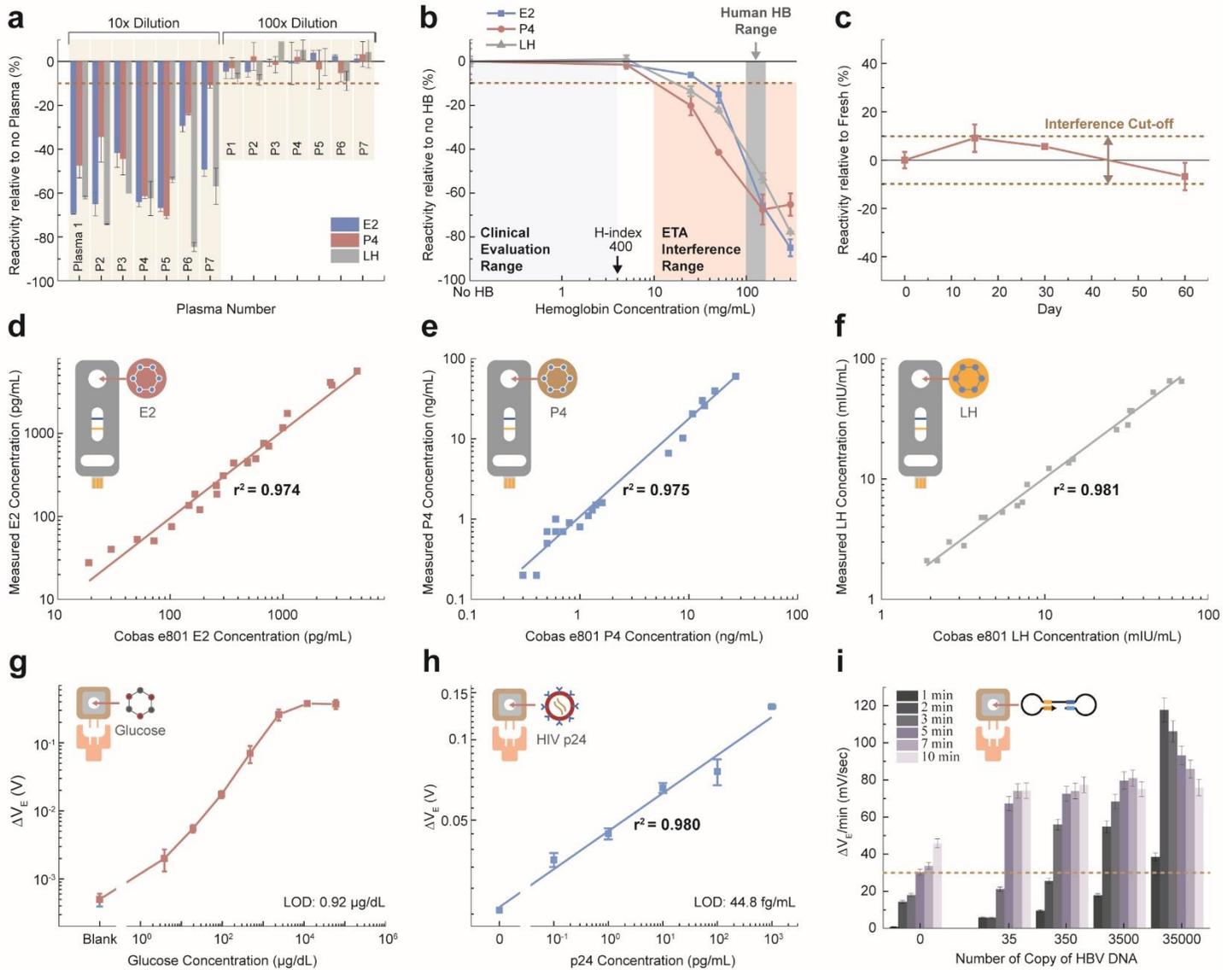

**Figure 4. Applicability of the ETA platform.** (a) Reactivity measured in human plasma samples for E2, P4, and LH after 10× and 100× dilution, with reactivity normalized to the no-plasma 1% BSA control. 100× dilution consistently brings responses within the cutoff, indicating effective mitigation of plasma-induced interference. (b) Reactivity as a function of HB concentration for E2, P4, and LH, normalized to the HB-free 1% BSA control. After 100× dilution, HB falls within the clinical evaluation range, where no measurable interference is observed. (c) Stability of LFIA-based ETA measurements. Reactivity of ChOx-dried LFIA strips measured over two months remains within <10% variation relative to freshly prepared strips. (d-f) Solid-state measurements using clinical plasma samples for clinically relevant analytes: (d) E2, (e) P4, and (f) LH, showing strong linear correlation with a Cobas e801 with accuracies of $r^2 = 0.974$ (E2), $r^2 = 0.975$ (P4), and $r^2 = 0.981$ (LH). Solution-state measurements include (g) a glucose enzymatic assay with an LOD of 0.92 µg/dL, (h) p24 detection via IMS workflow with an LOD of 44.8 fg/mL, and (i) HBV DNA detection using a LAMP workflow within 5 min.